\documentclass[a4paper,aps,prl,twocolumn,floatfix,showpacs,superscriptaddress,footnoteinbib]{revtex4-2}
\usepackage{blindtext}
\usepackage{graphics}
\usepackage{epsfig}
\usepackage{bbm}
\usepackage{times}
\usepackage{xcolor}   
\usepackage{amsfonts}
\usepackage{amsmath}

\usepackage{amssymb}
\usepackage{amsthm}
\usepackage{hyperref} 
\usepackage{wasysym}
\usepackage{bm} 
\usepackage{float}

\usepackage[normalem]{ulem}

\definecolor{mygreen}{rgb}{0.1, 0.6, 0.1}

\usepackage[utf8]{inputenc}

\begin{document}

\title{Synthetic Flat Bands, Hierarchical Topology, and Phase-Fluctuation-Insensitive Quantized \\ Transconductance in  Josephson Junctions}

\author{Subhadeep Chakraborty}
\email{subhadeep670@gmail.com, ORCID ID: 0000-0001-6651-0378}
\affiliation{Department of Physical Sciences, Indian Institute of Science Education and Research (IISER) Kolkata, Mohanpur 741246, India}

\author{Aabir Mukhopadhyay}
\email{aabir.phys@gmail.com, ORCID ID: 0000-0001-6465-2727}
\affiliation{Department of Condensed Matter Physics, Weizmann Institute of Science, Rehovot 7610001, Israel}
\affiliation{Department of Physical Sciences, Indian Institute of Science Education and Research (IISER) Kolkata, Mohanpur 741246, India}

\author{Udit Khanna}
\affiliation{Department of Physics, Bar-Ilan University, Ramat Gan 5290002, Israel}
\affiliation{Theoretical Physics Division, Physical Research Laboratory, Navrangpura, Ahmedabad 380009, India}
\author{Sourin Das}
\affiliation{Department of Physical Sciences, Indian Institute of Science Education and Research (IISER) Kolkata, Mohanpur 741246, India}

\begin{abstract}
We uncover hierarchy of topological phases within the synthetic Brillouin zone of a three-terminal Josephson junction’s (3-TJJ's) Bogoliubov–de Gennes spectrum. We demonstrate that the above-gap continuum realizes a Chern insulator phase with quantized monopole charges ($\pm1$), while the subgap Andreev bound states (ABS) are characterized by a quantized dipolar invariant. By breaking time-reversal symmetry at the junction, we induce synthetic flat bands that suppress DC Josephson currents across the entire phase-bias space. Furthermore, under voltage bias, the junction exhibits a robust quantization of the time-averaged transconductance that is reminiscent of a quantized Hall conductance plateau owing to the flat band limit and its  dipole phase. As a byproduct, the flat band produces a global “sweet plateau” of phase insensitivity, surpassing localized sweet spots of conventional superconducting qubits and enabling a robust architecture for symmetry-protected Andreev qubits.

\end{abstract}

\maketitle

\noindent
{\color{blue}{\textit {Introduction.}}} Flat bands with non-trivial topology\cite{PhysRevLett.103.206805, PhysRevLett.106.236802, PhysRevLett.106.236803, PhysRevLett.106.236804, PhysRevX.1.021014, PhysRevLett.107.146803, PhysRevB.62.R6065, bistritzer2011moire, sharpe2019emergent, serlin2020intrinsic} and topology-protected robust transport provide foundational pillars of modern quantum condensed matter physics, with significant potential for device applications and quantum information processing. While topologically non-trivial flat bands have been realized in solid-state platforms such as twisted bilayer graphene \cite{PhysRevResearch.2.033150, wu2021chern}, moiré superlattices \cite{waters2024topological, sarkar2025ideal}, and photonic lattices \cite{yang2017topological, yi2022strong}, each of these platforms presents distinct challenges, motivating the exploration of synthetic realizations.

In recent times, multi-terminal Josephson junctions (MTJJs) have emerged as a versatile platform for synthesizing tunable topological phases \cite{PhysRevB.92.155437, riwar2016multi, PhysRevB.95.075417, PhysRevLett.119.136807, PhysRevB.97.220505, PhysRevB.97.174518, PhysRevB.100.014514, PhysRevB.103.174504, PhysRevB.105.L241404, graziano2022selective, huang2022evidence, coraiola2023phase, gupta2023gate, PhysRevB.107.165301, PhysRevLett.134.156601, ram2025multiterminal}. By treating the independent superconducting phase differences as quasi-momenta within a ``synthetic Brillouin zone (s-BZ)',' these devices enable the simulation of high-dimensional topological matter in a controlled parameter space. The spectral architecture of a Josephson junction is fundamentally composite: the Bogoliubov spectrum naturally partitions into discrete, subgap Andreev bound states (ABS) and a continuous quasiparticle spectrum above and below the superconducting gap.

\begin{figure}[t]
	\centering
	\includegraphics[width=0.65\columnwidth]{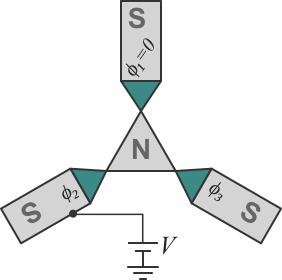}
	\caption{Schematic of a 3-TJJ realized with semiconducting nanowires and proximity-induced superconductivity. The independent phase biases $\phi_2$ and $\phi_3$ (relative to grounded $\phi_1$) act as synthetic quasimomenta, tunable via magnetic fluxes $\Phi_2=(\hbar \phi_2)/(2e)$ and $\Phi_3=(\hbar \phi_3)/(2e)$. A DC voltage bias $V$ is applied to the second terminal.}
	\label{3JJ_SetUp}
\end{figure}

In this Letter, we demonstrate that within the s-BZ, the entire Bogoliubov–de Gennes (BdG) spectrum, encompassing both the subgap Andreev bound states (ABS) and the continuum states, effectively approaches a flat-band limit (in the superconducting phase manifold). We reveal a hierarchical topology characterizing the many-body ground state where the continuum exhibits a monopolar Chern phase, whereas the subgap ABS bands manifest a dipolar phase (zero Chern number, but possessing a quantized dipole polarization). The signature of this flat Bogoliubov spectrum is imprinted on the transport response. Under a DC voltage bias applied to one terminal, the corresponding superconducting phase winds uniformly in time ($\phi \sim 2eVt/\hbar$), forcing the system to adiabatically traverse the s-BZ in the corresponding direction. This mechanism activates a topological pump: transconductance becomes quantized, governed by the integrated Berry curvature of the occupied sector. Crucially, we show that the hierarchical topology leads to a separation of transport contributions, where the quantized time-averaged conductance at the flat band limit arises entirely from the monopolar continuum, whereas the dipolar ABS contributes only to the oscillatory conductance when detuned from the flat band limit.

Flat Andreev levels emerge from destructive interference between electron and hole states — a superconducting analog of compact localized states(CLS)\cite{PhysRevB.95.115135,Leykam01012018}. These flat bands provide a robust pathway for engineering Andreev qubits\cite{PhysRevLett.90.087003,janvier2015andreev,hays2021andreev,schonenberger2024andreev} that are intrinsically decoupled from phase-bias fluctuations, extending the traditional ``sweet spot"\cite{ioffe1999environmentally} into a global ``sweet plateau". Consequently, our work unveils multi-terminal Josephson junctions as a platform where the coexistence of hierarchical topological phases within distinct spectral sectors is observed.

%%%%%%%%%%%%%%%%%%%%%%%%%%%%%%%%%%%%%%%%%%%%%%%%%%%%%%%%%%%%%%%%%%%
%%%%%%%%%%%%%%%%%%%%%%%%%%%%%%%%%%%%%%%%%%%%%%%%%%%%%%%%%%%%%%%%%%%

\noindent
{\color{blue}{ \textit{The Andreev bound-states.}}} We start by considering a normal scattering junction region of three one dimensional quantum wires characterized by a one-parameter family of real unitary scattering matrices\footnote{A $3 \times 3$ unitary scattering matrix has 9 real parameters, and 5 of them are removable by independent rephasings, $S \rightarrow e^{i A} S e^{i B}$, thus leaving the physically distinct parameter space to be four-dimensional. These phases $A=\text{diag}\{\alpha_1,\alpha_2,\alpha_3\}$, $B=\text{diag}\{\beta_1,\beta_2,\beta_3\}$ can be absorbed into the superconducting phases. Assuming the normal-region scattering matrix to be real restricts it to $O(3)$, which is three-dimensional. Hence this approximation eliminates only a single rephasing-invariant phase (parameter). Hence taking the $S$ matrix to be real does not lead to a significant compromise.}
\begin{align}\label{param_S}
    S=\exp \left[ \frac{i \zeta}{\sqrt{3}} (\lambda_2 + \lambda_5 + \lambda_7)\right],
\end{align}
where $\lambda_2$, $\lambda_5$ , and $\lambda_7$ are the known Gell-Mann matrices\cite{arfken2013mathematical}.
The three wires are assumed to host proximity-induced s-wave superconductivity (non-topological) and are connected through this scattering region, with macroscopic phase biases $(\phi_1, \phi_2, \phi_3)$ (see Fig. \ref{3JJ_SetUp}). Without loss of generality, we set $\phi_1=0$ as the global phase reference. In the short-junction limit, Andreev bound-state (ABS) energies $E$ of a Josephson junction (JJ) are determined by the determinant condition \cite{PhysRevLett.67.3836}
\begin{equation}
\det\!\left[1 - e^{-2i\arccos(E/\Delta)} Q\right]=0,
\label{eq:abs_condition}
\end{equation}
where, $Q=\,S\,e^{i\varphi}\,S^{*}\,e^{-i\varphi}$, $\Delta$ is real and represents the superconducting gap and $\varphi=\mathrm{diag} \{0, \phi_2, \phi_3 \}$.

\begin{figure}[t]
	\centering
	\includegraphics[width=\columnwidth]{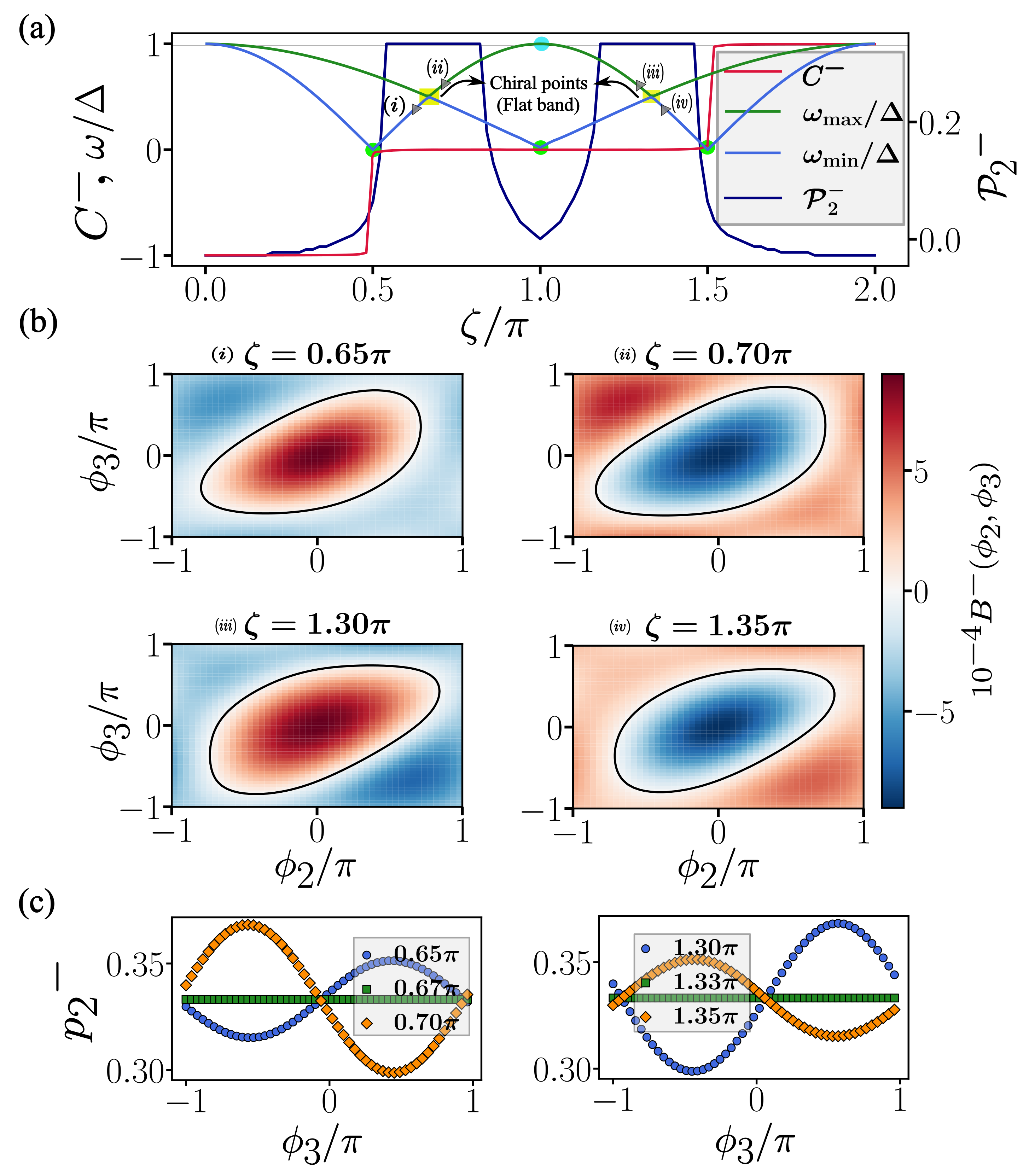}
	\caption{(a) The maximum and minimum gaps (in units of $\Delta$) in the ABS spectrum. Green circles indicate ABS band touchings at zero energy, the blue circle indicates the ABS bands touching the continuum, and yellow squares indicate the chiral points of the $S$-matrix. The Chern number ($C^-$) of the lowest occupied ABS band is plotted against $\zeta$. (b) Berry curvature of the lowest occupied ABS band for four parameter values around the chiral points: \textit{(i)} $\zeta=0.65\pi$, \textit{(ii)} $\zeta=0.7\pi$, \textit{(iii)} $\zeta=1.33\pi$, and \textit{(iv)} $\zeta=1.35\pi$ with zero Berry curvature contour separating positive and negative Berry curvature regions (c) The 1D dipole polarization ${p_2}^-$ is plotted against $\phi_3$ at and around the chiral points at different values of $\zeta$ depicted in the figure. }
    \label{analytical_fig}
\end{figure}

The matrix $S (S^*)$ describes the scattering matrices for electrons (holes) at the junction region. Using the above we obtain the plot for the maximum ($\omega_\mathrm{max}$) and minimum ($\omega_\mathrm{min}$) gap present in the ABS spectrum (see Fig. \ref{ABS_bands}, supplemental material) over the entire superconducting phase space $[-\pi, \pi)^2$ as depicted in Fig. \ref{analytical_fig} (a).

%%%%%%%%%%%%%%%%%%%%%%%%%%%%%%%%%%%%%%%%%%%%%%%%%%%%%%%%%%%%%%%%%%%
%%%%%%%%%%%%%%%%%%%%%%%%%%%%%%%%%%%%%%%%%%%%%%%%%%%%%%%%%%%%%%%%%%%

\noindent
{\color{blue}{\textit{The flat band.}}} We note that at $\zeta=2\pi/3 \approx 0.67\pi, 4\pi/3 \approx 1.33 \pi$, the maximum and minimum gap become equal, indicating a flat band nature of the ABS bands (see Fig. \ref{ABS_bands} of supplemental material for band dispersion). These points correspond to the chiral scattering matrix of the junction given by,
\begin{align}\label{chiral_S}
    S\left(\zeta=\frac{2\pi}{3}\right) = \begin{bmatrix}
        0 & 0 & 1 \\ -1 & 0 & 0 \\ 0 & -1 & 0
    \end{bmatrix} ; 
    S\left(\zeta=\frac{4\pi}{3}\right) = \begin{bmatrix}
        0 & -1 & 0 \\ 0 & 0 & -1 \\ 1 & 0 & 0
    \end{bmatrix}.
\end{align}
Chirality parameter $\delta G = (G_{i, i+1}^N-G_{i+1,i}^N)/(e^2/h)$\cite{PhysRevB.103.174504}, where $G_{i, j}^N=e^2/h ( \delta_{ij} - |S_{ij}|^2)$ is the normal state conductance between terminal-$i$ and $j$, quantifies the degree of non-reciprocity (breaking of time reversal symmetry) of the junction conductance. The matrices in Eq. \ref{chiral_S} correspond to $\delta G=\pm 1$, describing a pair of perfectly chiral junction which are time reversal partners of each other. Hence flat-band ABS bands, i.e., phase bias insensitive ABS, can be obtained by tuning the junction to one of the chiral point ($\zeta=2\pi/3 , 4\pi/3$). It is straightforward to establish this fact analytically by noting that the eigenvalues of $Q$ are independent of $\varphi$ at the chiral point and so are the solutions of Eq. \ref{eq:abs_condition}. 

Now we turn to the BdG continuum (states outside the gap) \cite{brouwer1997anomalous} whose density of states is given by,
\begin{align}\label{eq_cont_dos_1}
    \rho(E) &= - \frac{1}{\pi} \mathrm{Im.} \frac{d}{dE} \left[ \mathrm{ln} ~\mathrm{det} (1-S_AS_N) - \frac{1}{2} \mathrm{ln}~\mathrm{det} (S_AS_N)\right],
\end{align}
where the Andreev scattering matrix is given by $S_A = e^{-i \mathrm{arccos(E/\Delta)}} \tau_x e^{i\tau_z \varphi}$, with $\tau_i$ is the Pauli matrix acting in particle-hole space and $S_N$ is the normal scattering matrix given by $S$. When $S(\zeta)$ is tuned to the chiral point, the $\varphi$ dependence cancel out from the expression of $\rho(E)$ hence resulting in synthetic flat bands at all energies outside the gap (see FiG. \ref{cont_dos} of supplemental material). This implies that all eigenvalues of BdG Hamiltonian describing the system are independent of $\varphi$ at chiral point leading to perfect flat bands while the eigenfunction can still have $\varphi$ dependence leading to non-trivial band topology which is explored next.

%%%%%%%%%%%%%%%%%%%%%%%%%%%%%%%%%%%%%%%%%%%%%%%%%%%%%%%%%%%%%%%%%%%
%%%%%%%%%%%%%%%%%%%%%%%%%%%%%%%%%%%%%%%%%%%%%%%%%%%%%%%%%%%%%%%%%%%

\noindent
{\color{blue}{ \textit{ Topological phase of ABS bands.}}} The finite superconducting gap allows the spectrum to have two kinds of degeneracy (gap closing) point : ABS bands touch each other at zero energy for $\zeta=0.5\pi, \pi, 1.5\pi$, and, other case where ABS bands additionally touch the Bogoliubov continuum at $\zeta=\pi$. 
These two sectors- \textit{(i)} discrete ABS bands and \textit{(ii)} Bogoliubov continuum constitute the full Bogoliubov spectrum. In what follows, we will show that, away from degeneracy points, these two sectors manifest two distinct topological phases, even if they constitute the same many-body ground state.  While ABS band topology can be determined analytically, establishing the topology of the continuum sector requires a calculation using the many-body ground state and its excitation spectrum via numerical calculation of electrical current across the junction at finite voltage bias.

{\it \bf (i) The Chern phase:} Mapping the independent superconducting phases to synthetic crystal momenta ($(\phi_2, \phi_3) \to (k_x, k_y)$), we define a 2D s-BZ. Under this correspondence, the ABS play the role of Bloch bands of a lattice model, which provides direct access to the topological band theory toolbox, Berry curvature, Wilson loops, and other higher order multipole invariants \cite{PhysRevB.96.245115, PhysRevB.109.045402}. This mapping allows us to define the Chern number of the $\alpha$-th Andreev band in the s-BZ in complete analogy with a conventional band insulator given by,
\begin{equation}
C^\alpha = \frac{1}{2\pi}\!\!\int_{-\pi}^{\pi}\!\! d\phi_2
\int_{-\pi}^{\pi}\!\! d\phi_3 B^{(\alpha)}(\phi_2,\phi_3),
\label{eq:Chern}
\end{equation}
where the corresponding Berry curvature\cite{fukui2005chern, RevModPhys.82.1959, PhysRevB.104.085114} in s-BZ can be expressed as,
\begin{equation}
B^{(\alpha)} (\phi_2, \phi_3)=
\mathrm{Im.}\,\mathrm{Tr}\!\left[
P_\alpha
\big(\partial_{\phi_2}P_\alpha\,\partial_{\phi_3}P_\alpha
 - \partial_{\phi_3}P_\alpha\,\partial_{\phi_2}P_\alpha\big)
\right].
\label{eq:Fprojector}
\end{equation}
 Here $P_\alpha=|u_\alpha\rangle\langle u_\alpha|$ is the projector onto the $\alpha^{\text{th}}$ eigenstate of $Q$. Figure~\ref{analytical_fig}(a) shows the Chern number $(C^-)$ of the negative ABS band as a function of $\zeta$. It remains quantized between $\zeta=0$ and $\pi/2$. At $\pi/2$ the minimum gap $(\omega_{\min})$ closes and the system transitions to a trivial $(C^-=0)$ phase. A time-reversed counterpart of this transition occurs for $\zeta$ for $3\pi/2$. Clearly the flat bands lay outside the Chern phase of ABS. 
 
 To gain insight into the zero Chern phase hosting the flat bands, we evaluate the Berry curvature distribution in the Brillouin zone (BZ) around the chiral points (see Fig.~\ref{analytical_fig}(b)). The black contours denote zero-curvature lines that separate regions of equal and opposite integrated Berry curvature, indicating a finite dipole moment.

{\it \bf (ii) The dipole phase:} Motivated by the above, we evaluate the dipole polarization using Wilson loops \cite{PhysRevB.96.245115, PhysRevB.109.045402} defined on the synthetic toroidal manifold. We introduce dipole polarization as a geometric phase associated with polarization in synthetic space, providing deeper insight into the characteristics of the ABS bands beyond the monopole (Chern) invariant. 

 Assuming only the negative ABS band is occupied, we construct the Wilson loop along $\phi_2$ at fixed $\phi_3$ using the Abelian Berry connection ($\mathcal{A}_j^{(\alpha)}=
i\!\left\langle u_\alpha\middle|\partial_{\phi_j}u_\alpha\right\rangle$),
 \begin{equation}
W_2(\phi_3) =
\mathcal{P}\exp\!\left[
i\!\int_{-\pi}^{\pi}\!\!
d\phi_2\,\mathcal{A}^-_2(\phi_2,\phi_3)
\right]=e^{2\pi i p_2^-(\phi_3)},
\label{eq:Wilson}
\end{equation}
where $p_2^-(\phi_3) \in [-0.5, 0.5)$ denote the gauge invariant dipole polarizations for a ring along $\phi_2$ of the toroidal manifold at fixed $\phi_3$ values. The total dipole polarization in two dimensions is then obtained by taking a mean,
$\mathcal{P}_2^- = \mathrm{Mean}\{p_2^- (\phi_3)\}$. In Fig. \ref{analytical_fig} (c), we plot the magnitude of $p_2^-(\phi_3)$ as a function of $\phi_3$. The polarization remains flat at the chiral points but develops oscillations upon detuning, without changing the mean value. We note that the Wilson loop definition assumes periodic boundary conditions; consequently, the calculated $p_2^-$ captures magnitude rather than its sign. However, local sign inversions of the Berry curvature are distinctly marked by a reversal in the $\mathrm{sign}[\partial p_2^-/\partial \phi_3|_{\phi_3}]$ (see Fig. \ref{dipole_grad} of supplemental Material). Collectively, these results establish the zero Chern phase as a dipolar phase, where the flat ABS bands reside deep within this phase with a quantized dipole polarization $\mathcal{P}_2^-(\phi_3) \approx 0.33$ as depicted in Fig. \ref{analytical_fig} (a).

\noindent
{\color{blue}{\textit{Continuum topology.}}} While discussing band topology, the Chern number, central to topological classification, cannot be assigned to the ABS sector in isolation; rather, one must consider the topology of the entire filled BdG spectrum, including both subgap and continuum states. The total Chern number of the many-body ground state is manifested in the transconductance~\cite{riwar2016multi, PhysRevB.100.014521} of the Josephson Junction. Away from gap closing points, by subtracting the above calculated Chern number of ABS band from that total Chern number obtained  from transconductance, we obtain the Chern number of the Continuum.

Following Eriksson et. al. \cite{PhysRevB.95.075417}, Nowak et. al. \cite{PhysRevB.99.075416}, we proceed to calculate the transconductance by biasing the lead-$2$ with a voltage bias $V$ $(eV << \Delta)$ and measuring the current in the lead-$3$. The instantaneous current takes the form\cite{riwar2016multi, PhysRevB.100.014521},
\begin{align}\label{trans-1}
    I_3 (t) = \sum_{E<E_F} \left[\frac{2e}{\hbar} \frac{\partial E}{\partial \phi_3} -  2e \dot \phi_2 B_{32}\right] \left(n-\frac{1}{2}\right)
\end{align}
where $\dot \phi_2=2eV/\hbar$, and occupation number $n=1$ as we are measuring current contribution from levels below Fermi sea. Here $E$ and $B_{32}$ represents the energy and Berry curvature ($B_{32}=B(\phi_3, \phi_2)$) of the  negative energy states (ABS and continuum). The first term in Eq. \ref{trans-1} corresponds to the adiabatic current, while the second term is the first-order correction in the phase velocity, and encodes the quantum geometry of the entire occupied Bogoliubov spectrum.

The time-averaged (i.e., averaged over a full period of $\phi_2$) current depends only on the  contribution from quantum geometry of ABS and continuum and not on $E$ as it is periodic in $(\phi_1, \phi_2, \phi_3)$. Further, this contribution can be expressed as the rate of change of dipole polarization w.r.t. $\phi_3$ \cite{PhysRevB.96.245115, PhysRevB.109.045402} as,
\begin{align}\label{eq_dipole_curr}
    \bar I_3^g(\phi_3){\large/}\frac{2e^2 V}{h}  &= -\frac{1}{2\pi} \int_{-\pi}^{\pi} B_{32} d \phi_2 = \partial_{\phi_3} p^-_2 + \partial_{\phi_3} p^\mathrm{cont}_2
\end{align}
where $p_2^-$ and $p_2^\mathrm{cont}$ are 1D dipole polarizations of negative ABS band and continuum respectively. This allows for an interpretation of time-averaged current in terms of dipole moment. At chiral points, the ABS bands being flat with identically vanishing Berry curvature, the DC contribution of the instantaneous current, as well as the current contribution from ABS Berry curvature is zero. Note that continuum band are also flat at chiral point hence the entire contribution to current must come from the quantum geometry of the Bogoliubov continuum.

\begin{figure}[t]
	\centering
	\includegraphics[width=\columnwidth]{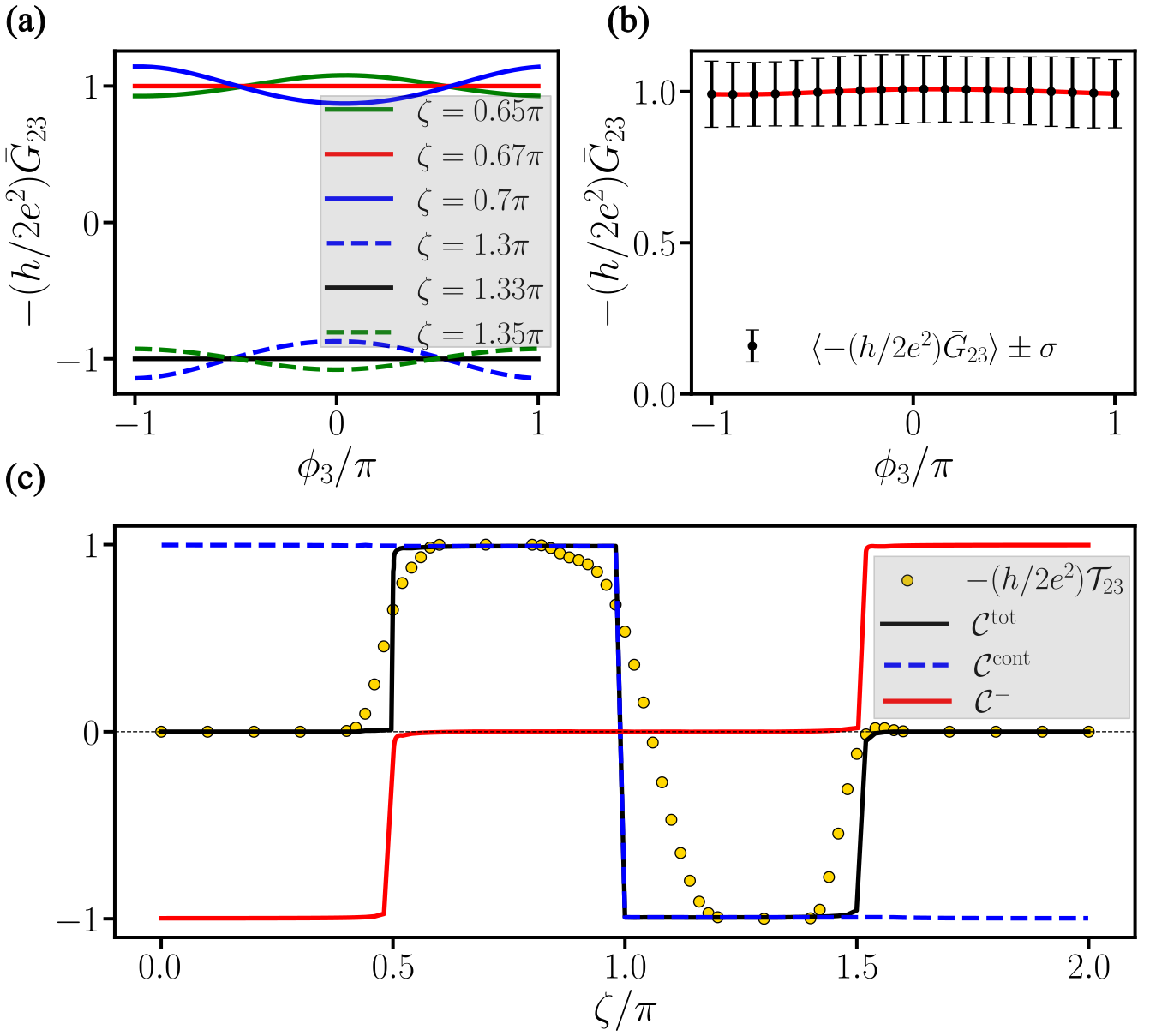}
	\caption{ (a) Time averaged conductance $ \bar G_{23}$ is plotted against $\phi_3$ at and around chiral points. (b) Time-averaged conductance at chiral point $\zeta \approx 0.67\pi$ is plotted in presence of random perturbation to the chiral scattering matrix (see Eq.~ \ref{perturbation}). (c) The transconductance, total Chern number, and Chern number of the continuum and ABS are shown. }
    \label{simul_fig}
\end{figure}

The numerically computed time averaged conductance $\bar G_{23}(\phi_3)=d \bar I_3(\phi_3)/dV$ is presented in Fig. \ref{simul_fig} (a). In the flat band limit, the time-averaged conductance becomes phase-insensitive (independent of $\phi_3$) and quantized, reminiscent of a quantized Hall conductance plateau. 
As discussed above, the subgap contribution in Eq.~(\ref{eq_dipole_curr}) vanishes due to the flat 1D ABS dipole polarization (see Fig.~\ref{analytical_fig}); consequently, the observed conductance quantization arises entirely from the continuum quantum geometry and can be attributed to the quantization of $\partial_{\phi_3} p^\mathrm{cont}_2$. Further more the integrated transconductance, $G_{23}=\int_{\pi}^{\pi}d\phi_3(dI_3(\phi_3/dV)$, calculated between the superconducting lead-$3$ and lead-$2$ of a multiterminal JJ is $\mathcal{T}_{3 2}=-(2e^2/h) C_{3 2}$ \cite{PhysRevB.100.014521}: where $C_{3 2}$ is the total Chern number of all occupied states \footnote{For a 3-TJJ, $C_{32}   =\int_{-\pi}^{\pi} \int_{-\pi}^{\pi} d \phi_3 d \phi_2 B(\phi_3, \phi_2) \\  = - \int_{-\pi}^{\pi} \int_{-\pi}^{pi} d \phi_3 d \phi_2 B(\phi_3, \phi_2) = - C_{23}  $} . As the Chern number of ABS in the flat band limit is zero, it leads us to the conclusion that the Chern number of the continuum is $C^\mathrm{cont}=C_{23}=\pm 1$ depending on sign of the Chirality parameter $\delta G$.

Thus, the flat-band limit exhibits a rich topological structure: {\it (i)} vanishing Chern number for the Andreev bound states, {\it (ii)}  continuum Chern number of $\pm 1$, {\it (iii)}  quantization of dipolar polarization of the Andreev sector, and {\it (iv)}  quantization of $\partial_{\phi_3} p^{\mathrm{cont}}_2$ associated with the continuum. This set of finding constitutes the central result of this Letter. As we move away from the chiral points, the quantum geometry of both ABS and continuum starts contributing, hence the time-averaged conductance becomes oscillatory in $\phi_3$. To check the robustness of this conductance quantization, we introduce all possible randomness perturbations to the chiral $S$-matrix, 
\begin{align}
    S' = \exp \left( \sum_{i=1}^{8} c_i \lambda_i + \frac{i}{\sqrt{3}} \frac{2\pi}{3} (\lambda_2+\lambda_5+\lambda_7) \right) 
    \label{perturbation}
\end{align}
where $c_i$ are randomly generated numbers within [$-0.1\pi, 0.1\pi$] following a uniform distribution. We plot $\bar G_{23}$ averaged over 100 distinct disorder realizations in Fig. \ref{simul_fig}(b), which reveals that the time-averaged conductance almost retains the plateau nature.

\noindent
{\color{blue}{\textit{ Non-trivial critical point at $\zeta=\pi.$}}} The sign reversal of the time-averaged conductance (transconductance) between the left and right chiral points is attributed to a global redistribution of monopole flux, triggered by simultaneous closing of minimum $\omega_{\mathrm{min}}$ and maximum $\omega_{\mathrm{max}}$ gap at different part of the synthetic Brillouin zone at $\zeta=\pi$, where reciprocity is restored. A cascading topological transfer drives this inversion. As can be seen from Fig. \ref{flux_inversion} of supplemental material, in the absolute neighborhood of gap-closing point $\zeta=\pi$, the negative energy ABS band transfers a monopole flux of -2 to the negative continuum (inside the zero Berry curvature contour); simultaneously, the positive energy ABS receives a charge of +2 from the negative ABS band (outside the zero Berry curvature contour). The positive ABS transfers +2 monopole flux to the positive continuum (inside zero Berry curvature contour). This redistribution flips the sign of the integrated Berry curvature within/outside the contour (see Fig. \ref{flux_inversion}),  effectively driving the negative continuum Chern number from +1 to -1 mediated by the ABS bands. Thus the ABS band retains a zero net Chern number; it acts as a topological bridge, facilitating the monopole flux transfer between the 
continuum bands which manifests itself in sign reversal in the quantized conductance plateau.

\noindent
{\color{blue}{\textit{ Discussion.}}} Our study highlights the coexistent monopolar and dipolar topology exhibited by different spectral fragments of the Bogoliubov spectrum (see Fig. \ref{sd_sch} of supplemental material). Owing to the exact chirality of the scattering matrix, the flat Andreev manifold emerges as immune to scattering phase perturbations or contact asymmetries. The experimental appeal of flat ABS bands lies in their reduced sensitivity to superconducting phase fluctuations. The concept of protecting a qubit from environmental noise by engineering its energy levels to be insensitive to the external control parameters at specific operating points has been a cornerstone of superconducting qubit engineering. A paradigmatic example is the quantronium qubit, where operation at a ``sweet spot" ensures the qubit transition frequency is first-order insensitive to the control parameter $\lambda$, that is $ \partial f_{01}/\partial \lambda=0$ \cite{ioffe1999environmentally}. 

\begin{figure}[t]
	\centering
	% Ensure the filename matches your local file
	\includegraphics[width=\columnwidth]{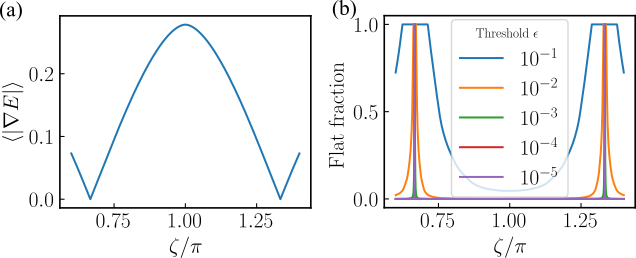}
	\caption{  Gradient of the ABS band energy with respect to superconducting phases averaged over the entire s-BZ is plotted in (a). Shown in (b) is the fraction of the entire superconducting phase manifold over which the flatness is achieved with different flatness thresholds $\epsilon$. }
	\label{qubit}
\end{figure}

Our work on the 3-TJJ presents a profound extension of this paradigm, moving from a point (``sweet spot") to a finite region of protection (``sweet plateau"). In the scattering parameter space, where the junction is described by chiral $S$-matrices, the Andreev bound state (ABS) bands become flat as a function of independent superconducting phase biases $\phi_2, \phi_3$. This renders absolute immunity to dephasing from phase (flux) noise. This is not merely a point of first-order insensitivity, rather a broad operating region where sensitivity up to any order identically vanishes. With $E$ being the negative ABS band energy, we calculate $|\nabla_\phi E|=|\sqrt{(\partial E/\partial \phi_2)^2+(\partial E/\partial \phi_3)^2}|$ averaged over entire s-BZ and plot it against the scattering matrix parameter $\zeta$ in Fig. \ref{qubit}(a). We consider a threshold value of parameter $\epsilon$, and define a quantity ``Flat fraction" to depict what fraction of the entire s-BZ shows $|\nabla_\phi E| < \epsilon$, which is plotted against $\zeta$ in Fig. \ref{qubit} (b). These two plots together establish the global flatness of the ABS band at and around chiral points, up to a small tolerance in the flatness.

The idea of preparing qubits using a MTJJ has early precedents. The D-Wave patent from 2001 \cite{amin2005quantum, amin2002multi}, the master's thesis \cite{jacobsen2023master} already envisioned that a qubit using MTJJ could be made. The thesis also reviews the broader context of protected qubits, such as the $\cos(2\phi)$ qubit, which aims for protection by having two minima per unit cell, a different mechanism that also results in a region of insensitivity at sweet spot.
More recent works on gate-tunable MTJJ have demonstrated the high level of control over these devices \cite{PhysRevB.101.054510, PRXQuantum.5.020340}.

Our findings however provides an elegant mechanism for achieving this protection through intrinsic chiral nature of $S$-matrices. The flat band arises due to perfect destructive interference of electron and hole trajectories, making perfectly canceling the superconducting phase dependence of ABS spectrum. This mechanism is a superconducting analog of compact localized states (CLS) in flat band lattice systems, where interference prevents dispersion.

A key consequence of the flat band condition is a fundamental design trade-off. Due to the energy and 1D dipole polarization of ABS bands being independent of superconducting phases, the Josephson current contribution from ABS band is zero, which makes the conventional methods of inductor-based qubit control and readout ineffective.

To overcome this, our paper insightfully suggests alternative readout strategies, such as coupling to high-$Q$ microwave resonator and probing the qubit via its state dependent quantum capacitance \cite{PhysRevLett.95.206807}. This aligns with the direction of modern circuit QED architectures such as ring resonators \cite{PhysRevApplied.16.024018}.

An additional direction is to exploit the spin degrees of freedom in such devices. The recent theory on the 3-terminal Andreev spin qubits \cite{55xn-c574} proposes using Andreev levels as a spin-1/2 system. In our flat-band scenario, this spin qubit would inherit the same phase-noise immunity, creating an exceptionally coherent spin qubit in an all-superconducting, voltage-controlled platform.
  
In contrast to current architectures for superconducting qubits—the vanguard of Noisy Intermediate-Scale Quantum (NISQ) computing—which rely heavily on “sweet-spot” engineering to mitigate dephasing, such protection is typically local, leaving qubits vulnerable to even modest drifts in control parameters. Our study demonstrates that breaking time-reversal symmetry in a 3-TJJ generates a flat synthetic Andreev band, resulting in a profound suppression of the Josephson current across the entire phase-bias parameter space.

This global phase insensitivity marks a paradigm shift from isolated points of protection to a robust “sweet plateau”, offering a blueprint for a new class of inherently decoherence-resistant memory qubits. Furthermore, under an applied bias voltage, the junction exhibits a quantized (time-averaged) transconductance that remains topologically robust against large phase fluctuations. By shifting the burden of coherence from active feedback to intrinsic Hamiltonian symmetries, our results provide a scalable route toward ultra-high-fidelity superconducting processors and long-lived quantum memories.

In summary, our study points out that the spectral segregation of the Bogoliubov spectrum of 3-TJJ suggests a novel utility: subgap modes function as passive, protected energy-storage elements or qubit isolators, while the continuum carrying the non-trivial Chern invariant mediates robust, quantized transport. This duality captures the essence of the hierarchical topological regime, establishing a single platform where topologically protected transport and geometric localization coexist.  Demonstrating a transition from a ``sweet spot" to a ``sweet plateau" for qubit protection via realizing a phase-insensitive Andreev level over a continuous region, our approach paves the way for qubits that are fundamentally immune to dephasing from their primary control parameters. The central challenge consequently shifts from protection to control—a problem the field is well-equipped to address using advanced circuit QED techniques. This positions the chiral 3-TJJ not merely as a simulator of topological matter, but as a promising building block for a new generation of ultra-coherent superconducting qubits.

\noindent
{\it \bf Acknowledgement:} We thank Julia Meyer for useful scientific correspondence. S.D. would like to acknowledge the financial support from Anusandhan National Research Foundation (ANRF) under the MATRICS scheme (Grant No. [ANRF/ARGM/2025/002511/TS)]);  Ministry of Education, Government of India under the SPARC program (Project Code: [SPARC/2025-2026/P4086]) and National Quantum Mission under Quantum Algorithms Technical Group (TPN No.: 136428). S.C. acknowledges the Council of Scientific and Industrial Research (CSIR), Government of India, for financial support in the form of a fellowship. We thank the computational facility (KEPLER) at the Department of Physics at IISER Kolkata. A.M. acknowledges The Weizmann School of Science, The Weizmann Institute of Science, Israel for financial support. S.C. acknowledges warm hospitality from the Institute of Physics (IoP) of Academia Sinica, Taiwan during the final stages of writing the draft.

%%%%%%%%%%%%%%%%%%%%%%%%%%%%%%%%%%%%%%%%%%%%%%%%%%%%%%%%%%%%%%%%%%%%%
\bibliographystyle{apsrev4-2}
\bibliography{References}

\clearpage

\pagebreak

\onecolumngrid
\pagestyle{empty}

\begin{center}

{\large \bf Supplemental Material: Synthetic Flat Bands, Hierarchical Topology, and Phase-Fluctuation-Insensitive Quantized  Transconductance in  Josephson Junctions}

{Subhadeep Chakraborty, Aabir Mukhopadhyay, Udit Khanna, and Sourin Das}

\end{center}

\section{Additional Plots: Analysis on topological phases of ABS bands and continuum}\label{App-A}

\begin{figure}[h]
	\centering
	\includegraphics[width=\columnwidth]{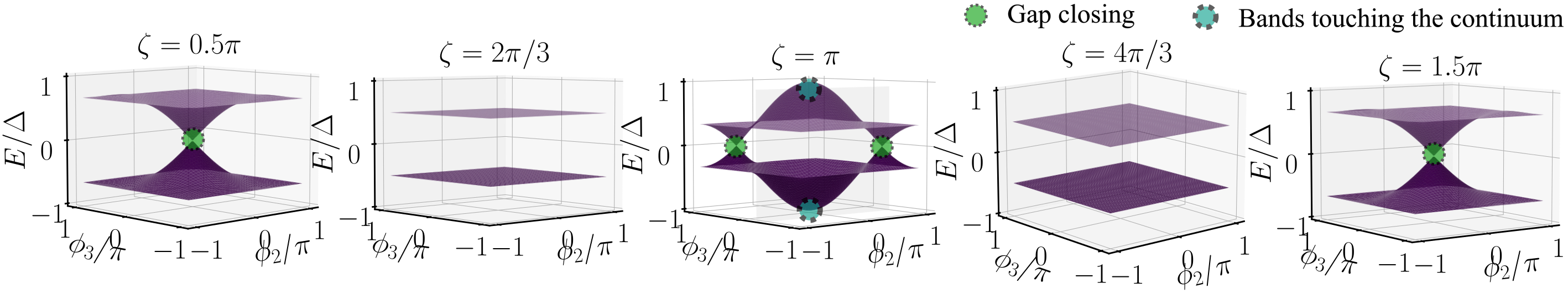}
	\caption{ ABS bands structure at different $\zeta$ values.
	}
	\label{ABS_bands}
\end{figure}

The ABS band structure obtained by solving Eq. \ref{eq:abs_condition} at different values of $\zeta$ is shown in Fig. \ref{ABS_bands}. The band structure plots show $\zeta=0.5\pi, \pi, 1.5\pi$ as ABS band gap closing, whereas $\zeta=\pi$ as the point where ABS bands touch the continuum. The ABS bands become flat at $\zeta=2\pi/3 \approx 0.67 \pi$ and $4\pi/3\approx 1.33\pi$.

\begin{figure}[h]
	\centering
	\includegraphics[width=0.6\columnwidth]{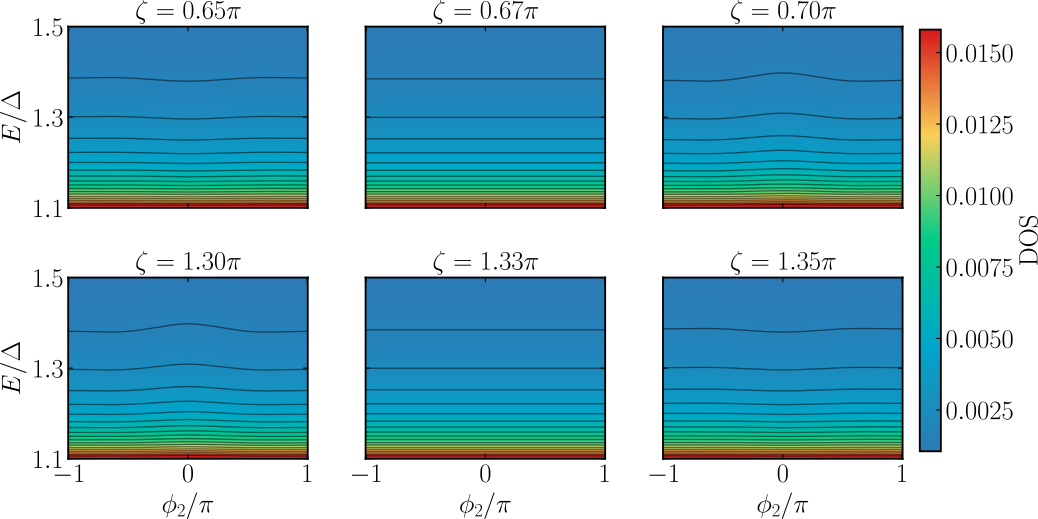}
	\caption{ Continuum density of states is plotted (heatmap and contour) at different $\zeta$ values. At chiral points $\zeta \approx 0.67\pi, 1.33\pi$, the continuum DOS is flat.
	}
	\label{cont_dos}
\end{figure}

\noindent
We plot the continuum DOS in Fig. \ref{cont_dos} along arbitrary line $\phi_2=-\phi_3$. The plots depict the continuum DOS is independent of the superconducting phase biases at chiral points.

\begin{figure}[h]
    \centering

    % Top row
    \begin{minipage}[t]{0.49\textwidth}
        \centering
        \includegraphics[width=\columnwidth]{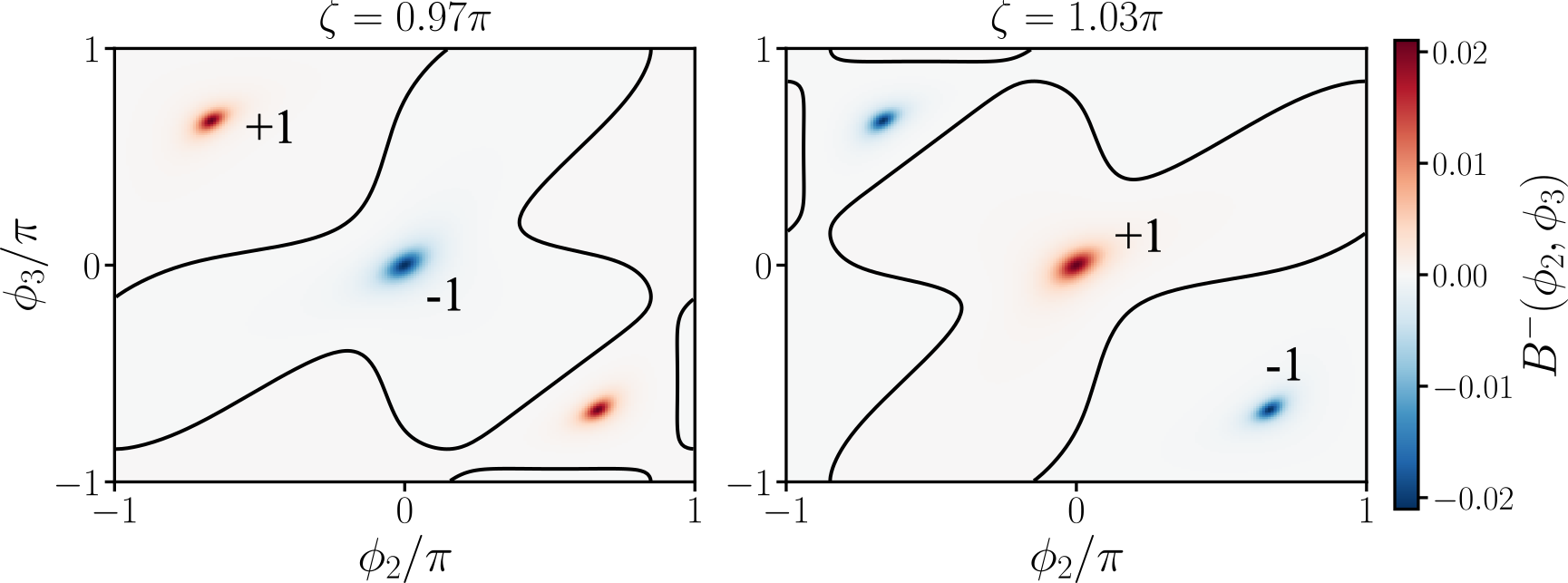}
        \caption{ Inversion of monopole flux distribution across the anomalous gap-closing point $\zeta=\pi$. The integrated Berry curvature within (outside) the zero Berry curvature contour is $-1 (+1)$, which flips sign across the gap-closing point $\zeta=\pi$.}
	\label{flux_inversion} 
    \end{minipage}
    \hfill
    \begin{minipage}[t]{0.49\textwidth}
        \centering
        \includegraphics[width=\columnwidth]{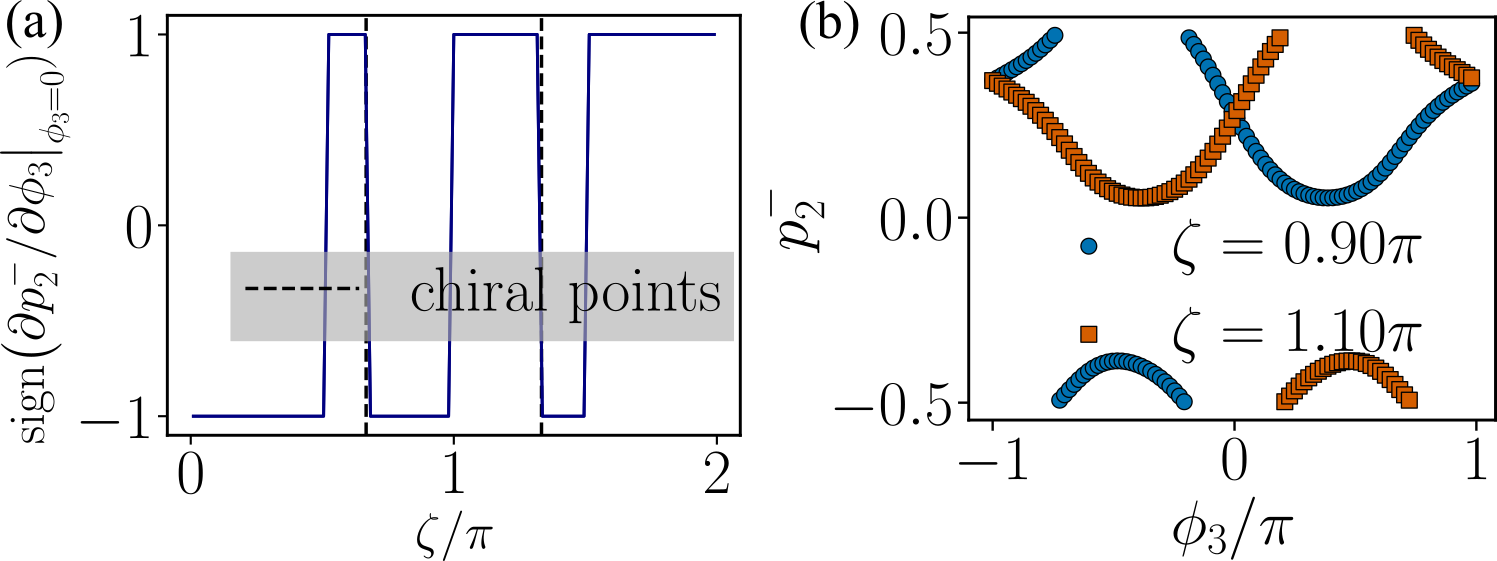}
        \caption{(a) Shows the inversion of the local dipole gradient (at $\phi_3=0$) in parameter ($\zeta$) space where the Berry curvature of the lowest ABS band inverts distribution. (b) Shows the profile of $p_2^-$ as a function of $\phi_3$ in the unquantized dipole phase.}
	\label{dipole_grad}  
    \end{minipage}

    \vspace{-0.1cm}
    
    % Bottom row
    \begin{minipage}[t]{\textwidth}
        \centering
	\includegraphics[width=0.75\columnwidth]{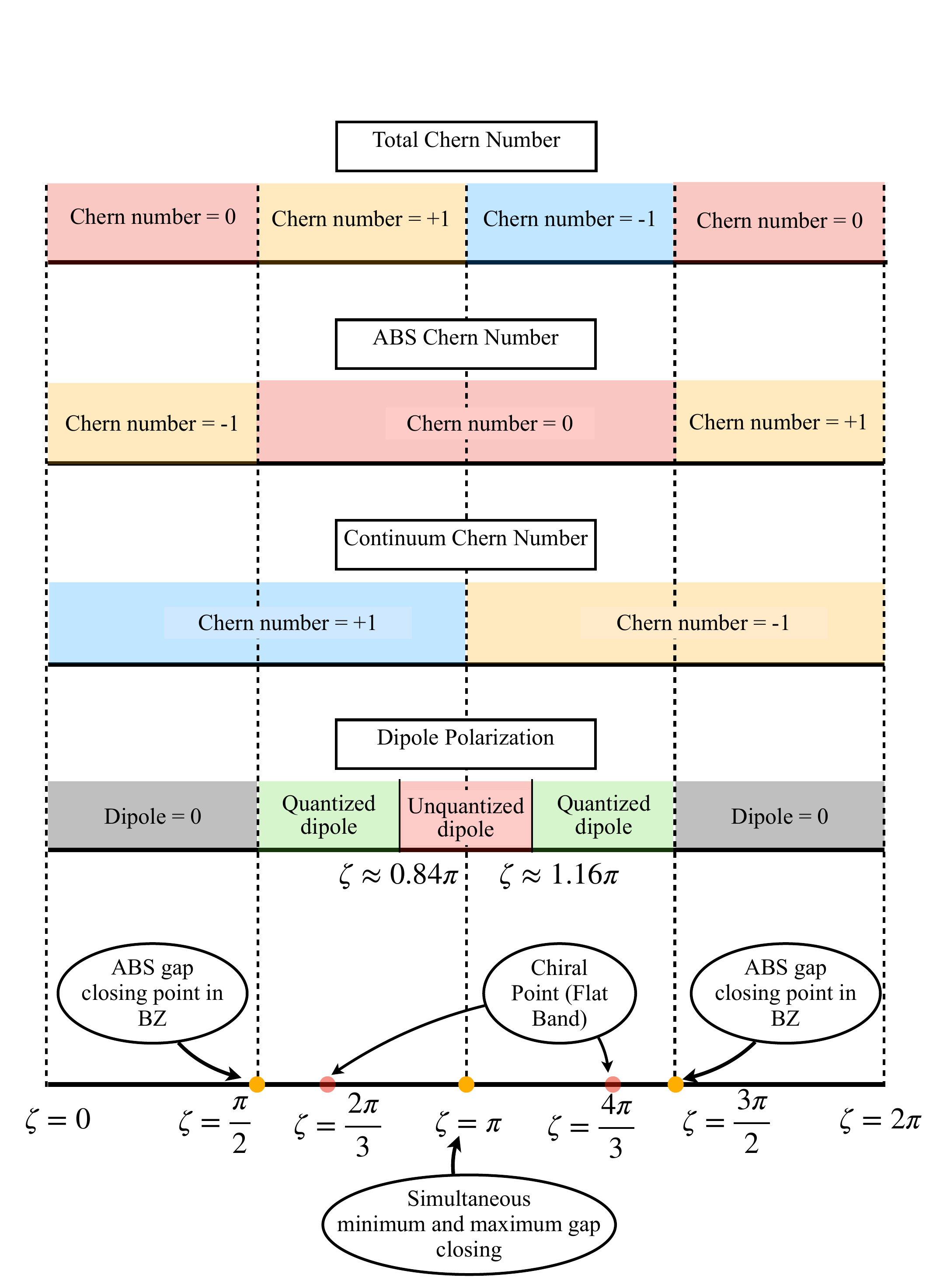}
	\caption{Schematic phase diagram showing gap-closings, Chern number transitions and corresponding ABS dipole polarization as a function of the scattering matrix parameter $\zeta$.}
	\label{sd_sch}
    \end{minipage}
\end{figure}

In Fig. \ref{flux_inversion}, we show the monopole flux (Berry curvature) distribution inversion in $\phi_2 - \phi_3$ plane across the anomalous gap closing point $\zeta=\pi$. This establishes the transfer of monopole flux between ABS bands and continuum.

In Fig. \ref{dipole_grad} (a), we plot the local dipole gradient ($\partial p_2^-/\partial \phi_3|_{\phi_3=0}$), which shows a sign reversal at each point in parameter space, where the Berry curvature distribution inverts. We plot the $p_2^-$ as a function of $\phi_3$ in the unquantized dipole phase in Fig. \ref{dipole_grad} (b). This shows jumps in the profile of $p_2^-$ due to wrapping it within $(-0.5, 0.5]$, leading to the unquantized nature of the total dipole polarization $\mathcal{P}_2^-$ in this phase.

These results, along with the analysis of the continuum Chern number (Fig. 3 of the main text) establishes a neat a phase diagram of the entire BdG spectrum of the 3-TJJ, shown in Fig. \ref{sd_sch}.
A schematic phase diagram showing the qualitative results in the parameter space is shown in FIG. \ref{sd_sch}.

\clearpage
\end{document}